\documentclass{ar-1col-S2O}
\usepackage{url}
\usepackage{rotating}
\usepackage{multicol, multirow}
\usepackage{array, booktabs}
\usepackage[bookmarks=true]{hyperref} %

\usepackage{amsmath, amssymb, amsthm}
\usepackage{caption}
\usepackage{array}  %
\usepackage{makecell}
\usepackage{subcaption}
\usepackage{balance} %
\usepackage{enumitem}
\usepackage{wrapfig}
\usepackage{tablefootnote}
\usepackage{colortbl}
\usepackage{bbm}
\usepackage{fontawesome5}  %
\usepackage{xcolor}

\usepackage[numbers]{natbib}
\jname{Annu. Rev. Control Robot. Auton. Syst.}
\jvol{AA}
\jyear{2024}

\newtheorem{theorem}{Theorem}
\newtheorem{corollary}{Corollary}[theorem]  %

\newtheorem{proposition}{Proposition}    
    
\newtheorem{definition}{Definition}

\usepackage{xparse}
\DeclareDocumentEnvironment{example}{o}{\summary[Example: #1]}{\endsummary}

\usepackage[nameinlink,
    capitalize,  %
    noabbrev,  %
]{cleveref}
\creflabelformat{equation}{#2#1#3}  %

\usepackage{ifthen}
\newboolean{include-notes}
\newboolean{mark-new}
\newboolean{include-remove}
\setboolean{include-notes}{false}
\setboolean{mark-new}{false}
\setboolean{include-remove}{false}

\usepackage[normalem]{ulem}
\definecolor{teal}{rgb}{0, 0.5, 0.5}
\definecolor{darkgreen}{RGB}{0,100,0}
\definecolor{porange}{RGB}{231, 117, 0}  %
\definecolor{turquoise}{RGB}{78, 190, 186}  %
\newcommand{\jaime}[1]{\ifthenelse{\boolean{include-notes}}{\textcolor{porange}{\textbf{Jaime:}\ #1}}{}}
\newcommand{\kc}[1]{\ifthenelse{\boolean{include-notes}}{\textcolor{blue}{\textbf{Kai:}\ #1}}{}}
\newcommand{\haimin}[1]{\ifthenelse{\boolean{include-notes}}{\textcolor{darkgreen}{\textbf{HH:}\ #1}}{}}
\newcommand{\remove}[1]{\ifthenelse{\boolean{include-remove}}{\textcolor{red}{\sout{#1}}}{}}
\newcommand{\new}[1]{\ifthenelse{\boolean{mark-new}}{\textcolor{teal}{#1}}{#1}}

\usepackage[colorinlistoftodos,prependcaption,textsize=footnotesize]{todonotes}
    \setuptodonotes{inline}

\usepackage{fontawesome5}  %

\newcommand{\cbf}{{h}}

\DeclareMathOperator*{\st}{{\mathop{\text{s.t.}}}}
\DeclareMathOperator*{\argmax}{{\mathop{\mathrm{argmax}}}}
\DeclareMathOperator*{\argmin}{{\mathop{\mathrm{argmin}}}}

\newcommand{\sgnDist}[1]{{s_{#1}}}
\newcommand{\indicator}{\mathbbm{1}}

\newcommand{\reals}{\mathbb{R}}

\newcommand{\integers}{\mathbb{Z}}
\newcommand{\compl}{\text{c}}  %

\renewcommand{\emptyset}{\O}  %

\newcommand{\prob}{{P}}
\newcommand{\probDens}{{p}}

\newcommand{\probd}{{\probDens_\dstb}}

\newcommand{\state}{{x}}

\newcommand{\ctrl}{{u}}

\newcommand{\dstb}{{d}}
\newcommand{\obsrv}{{y}}

\newcommand{\belief}{{b}}

\newcommand{\tplan}{{\tau}}

\newcommand{\xSet}{{\mathcal{X}}}
\newcommand{\cSet}{{\mathcal{U}}}
\newcommand{\dSet}{{\mathcal{D}}}

\newcommand{\nx}{{n_\state}}

\newcommand{\dyn}{{f}} %
\newcommand{\dyninfo}{\dyn^\info}

\newcommand{\dynCont}{\dyn^c}

\newcommand{\sense}{{h}}  %

\newcommand{\tcont}{{t}} %

\newcommand{\tdisc}{{k}} %

\newcommand{\khorizon}{{H}}

\newcommand{\infoSet}{{\mathcal{H}}}
\newcommand{\info}{{\eta}}

\newcommand{\outcome}{{J}}  %
\newcommand{\valFunc}{{V}}

\newcommand{\policy}{{\pi}}

\newcommand{\consFunc}{{g}}

\newcommand{\failure}{{\mathcal{F}}}
\newcommand{\constraint}{{\failure^\compl}}
\newcommand{\safeSet}{{\Omega}}
\newcommand{\safeSetMax}{{\safeSet^*}}

\newcommand{\reach}{{\hat\xSet}}

\newcommand{\cost}{{c}}

\newcommand{\policyCtrlOpt}{{\policy^{*}}}  %

\newcommand{\shield}{\text{\tiny{\faShield*}}}

\newcommand{\task}{{\text{task}}}
\newcommand{\safety}{{\text{s}}}

\newcommand{\monitor}{{\Delta}^\shield}
\newcommand{\shieldCriterion}{\monitor}%

\newcommand{\policyTask}{{\policy^\task}}
\newcommand{\fallback}{{\policy^\shield}}
\newcommand{\safetyFilter}{{\phi}}

\newcommand{\Hz}{~\text{Hz}}

\begin{document}

\markboth{Hsu et al.}{Safety Filters}

\title{The
Safety Filter:
\\A Unified View of Safety-Critical Control in Autonomous Systems
}

\author{Kai-Chieh Hsu, Haimin Hu, and Jaime F. Fisac
\affil{Department of Electrical and Computer Engineering, Princeton University, Princeton, New Jersey 08544, USA;
email: \{\href{mailto:kaichieh@princeton.edu}{kaichieh},\href{mailto:jfisac@princeton.edu}{haiminh},\href{mailto:jfisac@princeton.edu}{jfisac}\}@princeton.edu}
}

\begin{abstract}

Recent years have seen significant progress in the realm of robot autonomy, accompanied by the expanding reach of robotic technologies.
However, the emergence of new deployment domains brings unprecedented challenges in ensuring safe operation of these systems, which remains as crucial as ever.
While traditional model-based safe control methods struggle with generalizability and scalability, emerging data-driven approaches tend to lack well-understood guarantees,
which can result in unpredictable catastrophic failures.
Successful deployment of the next generation of autonomous robots will require integrating the strengths of both paradigms.
This article provides a review of safety filter approaches, highlighting important connections between existing techniques and proposing a unified technical framework
to understand, compare, and combine them.
The new unified view exposes a shared modular structure across a range of seemingly disparate safety filter classes and naturally suggests directions for future progress towards more scalable synthesis, robust monitoring, and efficient intervention.

\end{abstract}

\maketitle

\section{INTRODUCTION}

The rapid advancement of robotics and autonomous systems calls for rigorous safe control methods to ensure their reliable operation
across a wide range of environments. 
Recent years have witnessed significant progress in the development of computational tools and theoretical frameworks to address this imperative, ranging from model-based approaches such as control barrier functions~\cite{ames2014control}, model predictive control~\cite{wabersich2021predictive}, and reachability analysis~\cite{bansal2017hamilton}, to data-driven schemes featuring learned certificates~\cite{robey2020learning}, adversarial reinforcement learning~\cite{hsunguyen2023isaacs}, and self-supervised safety analysis~\cite{bansal2021deepreach}.
With these techniques having reached a certain degree of maturity, 
and new opportunities opened by ongoing advances in learning and optimization,
the time is ripe to take stock of the common principles and relative strengths of existing approaches
in a coherent review.
Unfortunately,
differences in terminology and background
between sub-communities
have resulted in a certain degree of fragmentation in the literature, with occasional reinvention of existing ideas under neighboring frameworks.
This review is \new{driven by} the conviction that connecting these approaches under a common theory will facilitate the transfer of insights between them, reducing the duplication of work and ultimately speeding up the much-needed progress in the field as a whole.
The present article aims to serve as a unifying review in three distinct ways.

\begin{enumerate}
    \item 
    \textbf{An introduction to the material}
    for students and researchers who are new to the area of safe decision-making and control, and who may be seeking a roadmap to navigate the rapidly growing volume of literature addressing this problem.
    \item
    \textbf{A distillation of \new{recent} advances}
    for researchers and practitioners in the field, who may be familiar with some but not all of the approaches to safety filtering and find themselves disconcerted (and perhaps exasperated) by the absence of clearly spelled out links between various coexisting methodologies and newly proposed ideas.
    \item 
    \textbf{A reflection on the state of the art} for the technical community as a whole, to prompt further discussion about the key challenges that the field will need to confront going forward, as well as the most promising avenues of research to address them.
\end{enumerate}

We emphasize that this article is by no means the first survey of safe control approaches. 
It is, however, unique in
providing a \emph{unified}
conceptual and theoretical framework
under which to identify common features and distill key ideas
from the plethora of recent research advances.
Surveys on specific safe control methodologies include Hamilton-Jacobi reachability~\cite{bansal2017hamilton,chen2018hj}, control barrier functions~\cite{ames2019control}, learning-based model predictive control~\cite{hewing2020learning,rosolia2018data}, stochastic model predictive control~\cite{mesbah2016stochastic,mesbah2018stochastic}, reachability analysis of neural networks~\cite{fazlyab2021introduction}, and safe control with learned certificates~\cite{Dawson2023safe}.
Brunke et al.~\cite{brunke2022safe} provides a comprehensive review of safe robot control in the context of reinforcement learning.

The present review is structured as follows:
\Cref{sec:formulation} discusses the general problem of safe decision-making under uncertainty,
introducing key concepts,
unifying terminology and notation from multiple areas,
and
motivating the idea of safety filters and safety-performance separation.
\Cref{sec:filters} reviews the most important classes of safety filters
(including some techniques not typically viewed as safety filters),
attending to their historical evolution
and
organizing them into a straightforward taxonomy.
\Cref{sec:unified} elucidates the common theoretical foundation that binds all safety filter approaches and further shows that \emph{any} provably correct safety filter can be analyzed 
based on what we refer to as the \emph{Universal Safety Filter Theorem}.
The review concludes with a discussion of the state of the art and future challenges in \Cref{sec:discussion}.

\section{SAFE DYNAMIC DECISION-MAKING} \label{sec:formulation}
To discuss and compare the many safety filter approaches, we first establish in this section a common technical language to describe them.

\begin{figure}[!t]
    \centering
    \includegraphics[width=1.0\textwidth]{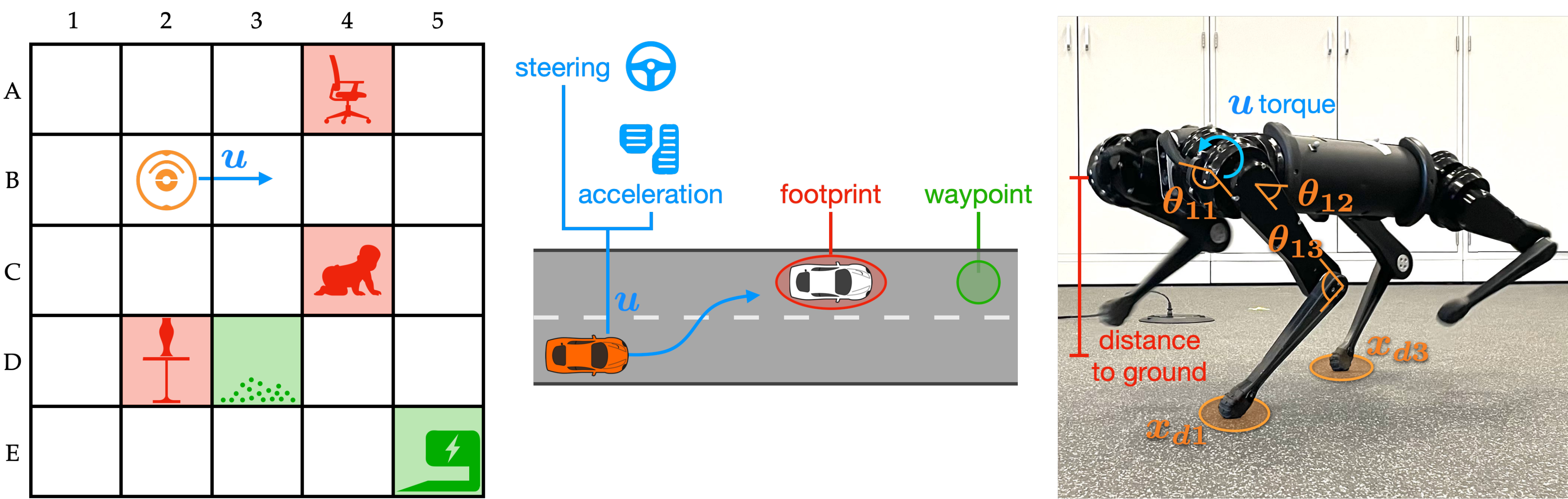}
    \caption{Examples of safety-critical autonomous systems. \emph{Left:} A robot vacuum cleaner navigating an environment represented as a 2-dimensional grid. \emph{Center:} An autonomous car driving on a highway. \emph{Right:} A quadrupedal robot walking in an indoor environment.}
    \label{fig:ex}
\end{figure}

\subsection{System Model} \label{subsec:dyn}
We consider an autonomous robot operating in a possibly dynamic environment.
We denote by $\state\in\xSet\subseteq\reals^\nx$ the \emph{state} of the overall system (i.e., the robot and its environment), a collection of variables that suffices to characterize and predict its behavior over time.
\subsubsection{Dynamic modeling and predictive uncertainty}
The evolution of the state $\state$
may be influenced by
certain variables that the robot can adjust instantly at each control cycle, termed
\emph{control inputs} (or \emph{actions}) $\ctrl\in\cSet\subseteq \reals^{n_\ctrl}$, where the set $\cSet$ usually captures actuator limits (e.g., motor torque).
The system is, in general, also subject to a \emph{disturbance input}, denoted by $\dstb\in\dSet$, which captures the model's \emph{predictive uncertainty}, including exogenous factors (e.g., wind), changes in operating conditions (e.g., motor wear), and overall modeling error.\footnote{%
    It is not that the function describing the evolution of the modeled $\state$ given $\ctrl$ over time cannot be computed exactly, but rather that such a ``true'' function may not exist at all.
    There may be two instances in which the same initial state and control signal result in different state trajectories.
}
Continuous-time dynamics,
${\tcont\in\reals}$,
naturally describe physical phenomena from first principles (e.g., Newton's law),
whereas
discrete-time models,
${\tdisc\in\integers}$,
match 
the finite \emph{control rate} at which robots compute and execute decisions.
Since most safe control frameworks
lend themselves to both paradigms,
we adopt the discrete-time formulation for brevity,
while cautioning that,
when reasoning with discrete-time models, it is important in practice to ensure that safety is satisfied \emph{in between} the time steps.\footnote{%
    Although not always addressed explicitly, the translation of discrete-time safety certificates into continuous-time guarantees has been studied using sampled-data system theory~\cite[e.g.,][]{mitchell2013safety}.
}
The system's evolution in discrete time can be described by
    \begin{equation} \label{eq:dyn_disc_uncert}
        \state_{\tdisc+1} = \dyn(\state_\tdisc, \ctrl_\tdisc, \dstb_\tdisc) \,. 
    \end{equation}
Note that 
an uncertain dynamical function
$\dyn: \xSet \times \cSet \times \dSet \to \xSet$ (and an associated set $\dSet$) 
always exists for any recorded system history $(\state_0,\ctrl_0,\state_1,\ctrl_1,\dots)$, such that \cref{eq:dyn_disc_uncert} is satisfied at each time $\tdisc$ for some value $\dstb_\tdisc\in\dSet$.\footnote{A trivial, extreme example is the ``agnostic'' model $\dyn(\state_\tdisc, \ctrl_\tdisc, \dstb_\tdisc):=\dstb_\tdisc$, with $\dstb_\tdisc\in\dSet \equiv \xSet$.}
\begin{marginnote}
    \entry{Control rate}{the frequency (typically expressed in$\Hz$) with which a robot can update its control input by executing a complete control cycle.
    }
\end{marginnote}

\subsubsection{Sensor modeling and perceptual uncertainty} \label{subsubsec:perc_uncert}

While the assumption of perfect state observability---always a simplification---is practically acceptable in many settings,
there are important cases in which it is not.
Significant perceptual uncertainty can result from, for example, sensor noise and occlusion.
The robot receives uncertain measurements modeled~as
\begin{equation}\label{eq:obsrv}
    \obsrv_\tdisc = \sense(\state_\tdisc, \ctrl_\tdisc, \dstb_\tdisc),
\end{equation}
analogous to the uncertain dynamics in~\cref{eq:dyn_disc_uncert}.
For conciseness, we let a single disturbance variable $\dstb_\tdisc\in\dSet$ encode uncertainty in both dynamics and perception.
When the state $\state_\tdisc$ cannot be directly observed, the robot's decisions instead rely on partial knowledge, represented in terms of an \emph{information state} $\info_\tdisc \in \infoSet$
that encodes the robot's initial state knowledge $\info_0$ and all subsequent observations $(\obsrv_1,\dots,\obsrv_\tdisc)$.
In many cases, $\info_\tdisc$ can be parameterized more succinctly without negatively impacting the robot's decision-making ability.
Under probabilistic uncertainty, for example, this \emph{sufficient statistic} takes the form of a \emph{belief} distribution $\belief_\tdisc\in\Delta(\xSet)$ over the current state.
Robust formulations, on the other hand, encode $\info_\tdisc$ as a set $\bar\xSet_\tdisc\subseteq\xSet$ of possible states.
The dynamics of $\info_\tdisc$ depend on~\cref{eq:dyn_disc_uncert,eq:obsrv}, but can be succinctly expressed as $\info_{\tdisc + 1} =  \dyninfo(\info_\tdisc, \ctrl_\tdisc, \dstb_\tdisc)$.

In practice, to reduce the overall state dimensionality, a number of methods
work with ``mixed observability'', i.e., a joint state $(\state_\tdisc, \info_\tdisc)$ combining perfectly observable state variables and an information state over uncertain ones ~\cite[e.g.,][]{mesbah2018stochastic,sunberg2022improving,hu2023active}.
Consistently with the bulk of the safe control literature,
we adopt the perfect state assumption
through much of the paper's exposition,
emphasizing the differences with imperfect information where relevant.
Safety filters for limited perception settings are discussed in detail in~\cref{subsec:perception_sf}.

\begin{marginnote}
    \entry{Information state}
    {a dynamic variable that encodes all relevant knowledge that is available for decision-making at a given time.}
    
    \entry{Belief state}
    {a conditional distribution over the system state $\belief_\tdisc := \probDens(\state_\tdisc \mid \mathbf{\obsrv}_\tdisc)$ where $\mathbf{\obsrv}_\tdisc$ is the observation history up to time $\tdisc$.}

    \entry{Sufficient statistic}
    {a compact representation of data that contains all the necessary information for decision-making.}
\end{marginnote}

\subsection{Safety Specifications} \label{subsec:safety_spec}
\new{Given a (possibly uncertain) system model, \emph{safety specifications} allow designers to encode critical requirements on system behavior and the terms under which these must be enforced.}

\subsubsection{Failure conditions} \label{subsubsec:state_cstr}
In many robotic applications, there exist potential failure conditions that are deemed unacceptable and must be avoided at all costs.
For example, autonomous vehicles should never drive off the highway,
robotic prostheses should never
force their users' joints past their range of motion,
and drones should never fall out of the sky.
These systems are called \emph{safety-critical}: maintaining safety is a hard requirement, not merely a desirable feature to be traded off with other performance metrics.
These catastrophic failure conditions can be concisely encoded by a \textit{failure set} $\failure\subseteq \xSet$ that the
state is forbidden from entering.
That is, the robot must always enforce the state constraint $\state\not\in\failure$, \new{or equivalently $\state \in \constraint$, where the complement $\constraint := \xSet \setminus \failure$ is called the \textit{constraint set}}.

\subsubsection{All-time safety and controlled-invariant sets} \label{subsubsec:inf_finite_guar}
While any given autonomous mission takes a finite amount of time, its duration usually exceeds the robot's explicit planning and prediction horizon.
As a result, we typically require the robot to ensure safety for \emph{all} future times, that is:
\begin{equation} \label{eq:safety_condition}
    \state_\tdisc \notin \failure, \quad
    \forall \tdisc \geq 0
    \,.
\end{equation}
Naturally, the robot cannot explicitly check an infinite sequence of time steps.
Instead, all-time safety can be implicitly guaranteed through the property of \emph{set invariance}. 
For now, we focus on the fully observable case ($\obsrv \equiv \state$).
A set $\safeSet \subseteq \xSet$ is 
\emph{controlled-invariant} under the dynamics in~\cref{eq:dyn_disc_uncert} if there exists a control policy $\policy:\xSet\to\cSet$ such that
\begin{equation} \label{eq:ctrl-inv-set}
    \forall \state\in\safeSet, \, \dyn \big(\state,\policy(\state),0\big) \in \safeSet \, .
\end{equation}
In other words, the robot can prevent the system from \emph{ever} leaving the set by continually applying policy $\policy$.
This deterministic definition ($d=0$) can readily be extended to \emph{robust} or \emph{probabilistic} controlled-invariant sets
in the presence of disturbances.\footnote{%
    Similarly, a controlled-invariant information set~$\safeSet\subseteq\infoSet$ may be defined when $\obsrv\not\equiv\state$.
}
If the controlled-invariant set~$\safeSet$ is disjoint from the failure set, $\safeSet \,\cap\, \failure = \emptyset$, then
it is called a \emph{safe set} \new{(see the asides titled Safe Set vs. Constraint Set and Safety vs. Stability)}.

\begin{marginnote}
    \entry{Robust controlled invariant set}
    {a set $\safeSet\subset\xSet$ is robustly controlled-invariant under bounded uncertainty $\dstb\in\dSet$
    if there is a control policy $\policy: \xSet \to \cSet$ that enforces $\dyn \big(\state,\policy(\state),\dstb\big) \in \safeSet$ for any $\state\in\safeSet$ and for \emph{all} values $\dstb\in\dSet$.}
    \entry{Probabilistic controlled invariant set}
    {a set $\safeSet\subset\xSet$ is probabilistically controlled-invariant with confidence level $\delta\in[0,1]$ under uncertainty $\dstb_\tdisc\sim\probd$
    if there is a control policy $\policy: \xSet \to \cSet$ that enforces $\prob( \state_\tdisc \!\in\! \safeSet, \forall \tdisc \!\ge\! 0) \!\!\ge\!\! \delta$ for any $\state_0\in\safeSet$.}
\end{marginnote}

\begin{textbox}[h!]\section{Safe Set vs. Constraint Set}
Note that the notion of safe set~$\safeSet$ is not equivalent to the state constraint set $\failure^\compl$,
and in fact the set $\failure^\compl$ is not in general a safe set.
There usually exist states that are not explicit failure conditions but from which the robot cannot prevent a future safety violation, and they are therefore \emph{unsafe states}.
For example, a drone not currently in collision is still unsafe if it is flying towards an obstacle too fast to avoid it.
\end{textbox}

\begin{textbox}[h!]\section{Safety vs. Stability}
While related, safety and stability are distinct properties, and neither is in general sufficient for the other (a system may be safe but not stable, or stable but not safe).
Certain stability methods, such as Lyapunov analysis, allow establishing a forward-invariant \emph{region of attraction} around an equilibrium state; if this region happens to be disjoint from the failure set~$\failure$, it is by definition a safe set~$\safeSet$.
Due to this, stability analysis tools are often used to construct sufficient conditions for all-time safety.
\end{textbox}

In most cases, finding \emph{some} safe set~$\safeSet$ is not 
particularly difficult;
however finding one \emph{large enough} to be of practical utility is less straightforward.
In a static environment, for example, the set of collision-free at-rest robot configurations is a safe set for collision avoidance purposes:
the robot can remain in this set, and thereby collision free, by never moving;
of course, some hard-to-please customers may return the robot on the basis that it doesn't do anything.
Less extreme (but similarly impairing) examples were common in the early years of social robot navigation
\citep[see the ``freezing robot'' problem,][]{trautman2010unfreezing}
and even autonomous driving technology,
where prototype vehicles would
comically
settle for the off-ramp after failing to secure a safe merge onto a moderately busy highway~\cite{reuters2018waymo}.
The need to ensure safety without needlessly hindering the robot's operation (further discussed in \cref{subsec:obj})
motivates the notions of \emph{maximal safe set}~$\safeSet^*$
and \emph{optimal safety policy}~$\policy^*$.
That is, from what states is it \emph{possible} for the robot to maintain safety,
and how?

\subsubsection{Safety in the face of uncertainty: robust vs. probabilistic guarantees} \label{subsubsec:rob_prob_gua}
Robust safety formulations aim to find a control strategy for the robot that can guarantee safety for all possible realizations of the uncertainty within a bounded (although usually infinite) set.
That is, the robot's policy $\policy: \xSet \to \cSet$ should enforce the robust safety condition
\begin{equation} \label{eq:safety_condition_robust}
    \state_\tdisc \notin \failure, \quad
    \forall \tdisc \geq 0, \;
    \forall \dstb_0, \dstb_1, \dots \dstb_\tdisc \in \dSet
    \,.
\end{equation}
In contrast, probabilistic safety formulations aim to satisfy a \emph{chance constraint} under a given distribution of uncertainty realizations (typically $\dstb_\tdisc \sim \probd$ i.i.d.).
Policy~$\policy$ then enforces
\begin{equation} \label{eq:safety_condition_chance}
    \prob \big(
        \state_{\tdisc} \notin \failure,
        \forall \tdisc 
        \ge 0
    \big) \geq 1 - \delta \, .
\end{equation}
In other cases, the chance constraint is instead imposed at each time step (marginally) or for a finite horizon $\tdisc \in \{ 0, 1, \cdots, \khorizon \}$.
Note that the chance constraint of~\cref{eq:safety_condition_chance} becomes equivalent to the robust constraint of \cref{eq:safety_condition_robust} when $\delta = 0$,
as long as the support of the distribution $\probd$ is $\dSet$.
Conversely, for $\delta>0$, safety may be compromised even when all modeling assumptions hold, since catastrophic failures $\state_{\tdisc} \in \failure$ are allowed to happen with a small non-zero probability.

\begin{marginnote}
    \entry{Safe set}
    {a region in the state space from which the robot can ensure safety at all future times.}
    \entry{Maximal safe set}
    {the \new{(unique)} set of all states from which it is possible for \emph{some} control policy to maintain safety for all time.}
    \entry{Optimal safety policy}
    {a control policy that enforces all-time safety from \emph{all}
    states from which maintaining safety is in fact possible.}
\end{marginnote}

Probabilistic safety guarantees are appropriate in settings with high uncertainty and limited stakes:
a robot vacuum cleaner should not damage valuables or bump into toddlers (\cref{fig:ex}, left), but occasional failures may be acceptable as long as they are sufficiently rare.
This contrasts with high-stakes domains like autonomous driving, where near-zero
failure probabilities can cause important practical issues.
Many types of safety-critical events are unlikely due to their extreme nature
and, as a result, may fall through the cracks in a probabilistic assessment,
even when their handling may be 
crucial to the system's viability.\footnote{%
    The fact that a safety-critical system may fail \emph{systematically} in certain low-probability conditions and still be deemed ``statistically safe''
    underpins the ongoing regulatory investigations and public controversy around the safety standards of some automated driving systems~\cite{national2020collision,crash2022NYT}.
}
The difficulty in achieving acceptable safety levels through statistics-informed design and evaluation
is a notoriously long-standing challenge in the autonomous driving field, where it is commonly referred to as the ``long tail''~\cite{makansi2021exposing}.

\subsubsection{Operational design domain: qualifying safety} \label{subsubsec:odd}
A commonly voiced objection against technical approaches seeking rigorous safety guarantees is that they aim to ensure ``absolute safety'',
thereby setting an impossible standard.
This is a fundamental misinterpretation of what safety guarantees \emph{are}.
The wide range of principled safety frameworks---and safety filters in particular---can best be viewed as establishing
\textit{conditions} under which a particular system will not experience a \emph{specified class} of catastrophic failures.
In the last decade, the autonomous driving community has led significant efforts to formalize the notion of an engineering system's \emph{operational design domain} (ODD),
a clearly delineated set of operating conditions under which it can be expected---and required---to function correctly and safely~\cite{sae3061,RR-2662}.
These may, for example, include ``environmental, geographical, and time-of-day restrictions, and/or the requisite presence or absence of certain traffic or roadway characteristics''~\cite{sae3061}.
\begin{marginnote}
    \entry{Operational design domain}
    {the set of conditions in which an engineering system, particularly an automated one, has been specifically designed to operate.}
\end{marginnote}

More broadly, we can apply this idea to any autonomous robotic system: what \emph{restrictions and assumptions} need to be imposed on its operating conditions (where and how it is deployed, the range of terrain and weather it may encounter, the behavior of nearby agents) in order to ensure safety?
The form of these assumptions is often tied to the sought guarantees: probabilistic safety assurances follow most naturally from statistical assumptions,
while robust guarantees are more readily derived from assumed bounds on uncertain variables.
In this sense, the modeling of uncertainty as either $\dstb_\tdisc\sim\probd$ or $\dstb_\tdisc\in\dSet$, with the associated prediction and perception models, can be viewed as establishing an operational design domain for the safety guarantees offered by different approaches.
Given their practical importance, the review in~\cref{sec:filters} pays special attention to the assumptions and guarantees associated to different classes of safety filters.

\subsection{Safety Filters: Reconciling Safety and Performance}
\label{subsec:obj}

\begin{marginnote}
    \entry{Performance objective}
    {quantification of the robot's success at a given task,
    typically defined as a combination of metrics
    accounting for factors like
    completion time, energy usage, and human user comfort and satisfaction.}
\end{marginnote}

Naturally, robotic systems should not only avoid reaching catastrophic failure states (this, by itself, can be accomplished by not turning them on),
but they should carry out their assigned tasks reliably and efficiently.
Constraining a robotic system to remain in an overly restrictive safe set can render it impractically inefficient at its intended function, or even unable to fulfill it.
The desirable robot operation is therefore one that achieves the highest possible task performance while on the other hand ensuring that safety will be maintained at all times
(with the required type of guarantee and under the assumptions comprising its operational design domain).
Formally, 
the overall goal is to find a robot control policy $\policy \colon \xSet \to \cSet$ to optimize a prescribed performance objective---which depends on
the state and control over time---subject to the appropriate safety condition:
\begin{subequations}
    \label{eq:soc}
\begin{eqnarray}
    \outcome^*(\state_0) = & \displaystyle \
    \min_{\policy} \ &
    \outcome(\state_0, \policy)
    \label{subeq:soc_obj}
    \\
    & \st &
    \text{safety condition (\Cref{eq:safety_condition_robust} or \Cref{eq:safety_condition_chance})}.
    \label{subeq:soc_safety}
\end{eqnarray}
\end{subequations}
\new{
\looseness=-1
Importantly, the uncertainty characterization
in the performance objective~$\outcome$ is, in general, distinct from 
the one
in the safety condition: for example, Objective \ref{subeq:soc_obj} may encode an \textit{expected} cost with $\bar\dstb_\tdisc\sim\bar\probDens_\dstb$, while Constraint \ref{subeq:soc_safety} may consider \textit{robust} safety under $\dstb_\tdisc\in\dSet$.
Finding a suitable~$\policy$ to jointly achieve performance and safety
can prove challenging, and cumbersome to repeat for each target task.
Fortunately, the problem can be broken down.
}

A \textit{safety filter} is an automatic process that monitors the operation of an autonomous system at runtime and intervenes, when deemed necessary, by modifying its intended control action 
in order to prevent a potential catastrophic failure.
The controlled invariance property of the maximal safe set  $\safeSet^*$ implies the existence of a ``perfect'' or ``least-restrictive'' safety filter that enforces the safety requirement while only overriding control actions that would cause the system to immediately leave $\safeSet^*$.
\begin{proposition}[Perfect Safety Filter] \label{prop:perfect_filter}
Consider a system with dynamics as in~\cref{eq:dyn_disc_uncert} 
under uncertainty $\dstb\in\dSet$
and safety requirement as in~\cref{eq:safety_condition_robust}.
For each state~$\state\in\xSet$, let $\cSet^\safety(\state)\subseteq\cSet$ contain all control actions that
\new{robustly
keep the 
next
system state inside} the maximal safe set~$\safeSet^*$.
Then, there exists a safety filter $\safetyFilter \colon \xSet \times \cSet \to \cSet$ such that
1) 
$\safetyFilter(\state, \ctrl) \in \cSet^\safety(\state)$ wherever $\cSet^\safety(\state)$ is not empty,
2) 
$\safetyFilter(\state, \ctrl) = \ctrl$ whenever $\ctrl \in \cSet^\safety(\state)$,
and
3)
the filtered system satisfies the original safety requirement.
Additionally, the safe decision problem in~\cref{eq:soc} can be transformed into a safety-agnostic decision problem through~$\safetyFilter$:%
\begin{subequations}
\label{eq:filter_soc}
\begin{eqnarray}
    \outcome^*(\state_0) = & \displaystyle \ \min_{\policyTask} \ &
    \outcome(\state_0, \policyTask)
    \label{subeq:filter_soc_obj}
    \\
    & \st 
    & \ctrl_\tdisc = \safetyFilter\left( {\state}_\tdisc, \policyTask ({\state}_\tdisc) \right).
    \label{subeq:filter_soc_safety}
\end{eqnarray}
\end{subequations}
\end{proposition}
\looseness=-1
The above result means that it is possible to design a safety filter for a robotic system with no knowledge or consideration of the particular task objectives,
only requiring the filter to be as permissive as possible while preventing any safety violations.
In turn, the task-oriented control policy can then be relieved of the burden of maintaining safety, since it can be viewed as acting on a modified system shielded by the safety filter.
From the task policy's perspective, this filtered system is \emph{incapable} of entering the failure set, regardless of the commanded control actions.
Because only unsafe actions are preempted by the filter, the robot is still able to achieve the optimal viable task performance.
This decoupling between safety and performance serves as a sort of separation principle, enabling the independent design of safety-agnostic task policies and task-agnostic safety filters \new{(see the Safe Learning aside)}.

\begin{textbox}[h!]\section{Safe Learning}
\new{In recent years, the safety-performance separation afforded by safety filters has proven particularly useful in enabling \emph{safe learning} schemes for robots and other safety-critical autonomous systems. Indeed, this goal has driven the development of a variety of ``safe reinforcement learning'' approaches whose core element is a suitable type of safety filter $\safetyFilter$ that allows a conventional (safety-agnostic) learning algorithm to freely explore and iteratively update $\policyTask$ without damaging the physical system or its environment.}
\end{textbox}

Unfortunately, such a least-restrictive safety filter cannot be tractably obtained for most robotic systems of practical interest.
Instead, modern safety filters seek diverse tradeoffs between permissiveness, robustness, generality, and computational scalability.
In the next section,
we review
the most prominent classes of safety filter schemes,
which, despite their algorithmic differences, share a core structure and important characteristics that make it possible to compare and categorize them.
Then, we establish in~\cref{sec:unified} a unified theoretical framework that distills this structure to 
\new{reveal a universal working principle}.

\section{SAFETY FILTERS} \label{sec:filters}

\begin{figure}
    \centering
    \includegraphics[width=\textwidth]{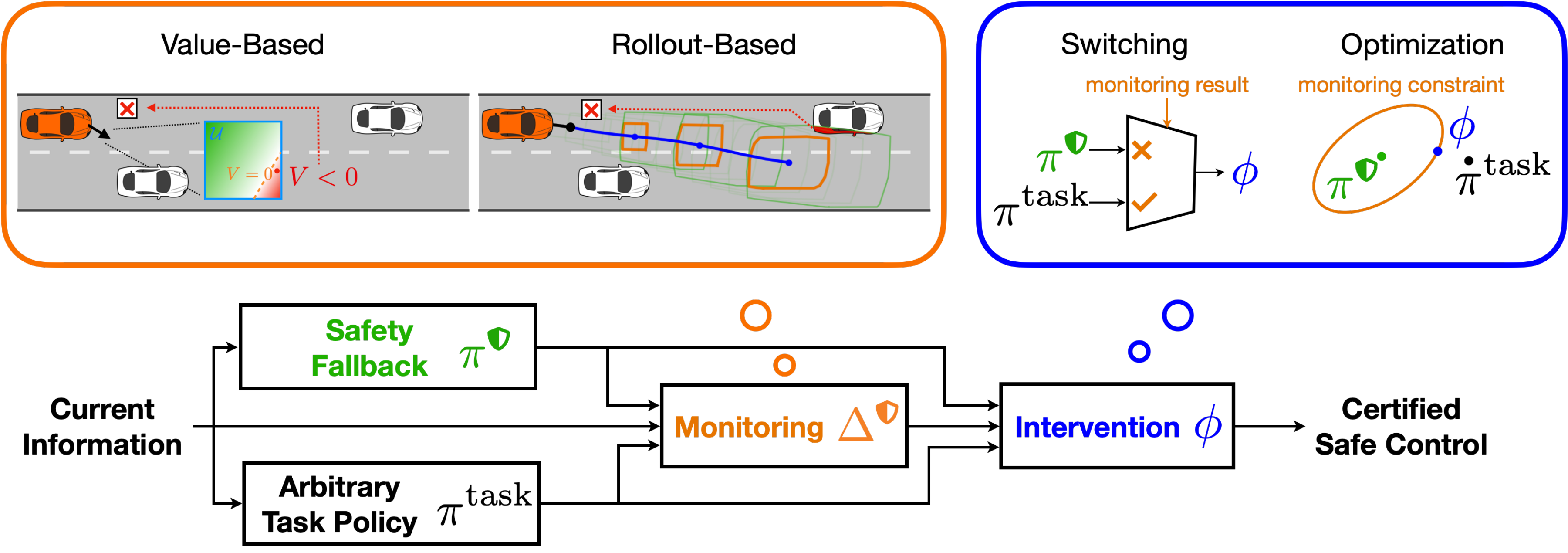}
    \caption{A safety filter consists of
    1) a safety monitor $\shieldCriterion$, which determines whether the control proposed by a task policy $\policyTask$ is safe, 
    2) an intervention scheme, which modifies the candidate control as needed to satisfy the safety requirement, and 
    3) a safety-oriented fallback policy $\fallback$, which informs monitoring and intervention.
    Many monitoring approaches can be catergorized as \emph{value-based} (evaluating a function that encodes whether a state or action is safe) or \emph{rollout-based} (computing an explicit prediction of the system's potential evolution).
    Common intervention approaches include \emph{switching} between the task policy and the fallback strategy and \emph{optimizing} the control to accommodate task performance criteria within the confines of safety.
    Figure adapted from Hsu et al.~\citep{hsunguyen2023isaacs}.}
    \label{fig:safety_filter_block}
\end{figure}

\looseness=-1
The runtime operation of every safety filter can be conceptualized as two complementary functions: \emph{monitoring} and \emph{intervention}.
The safety filter continually monitors the robot's proposed course of action to determine whether (or to what degree) it poses a safety risk.
Accordingly, the filter may intervene by modulating or completely overriding the robot's candidate control input to ensure that safety is maintained.
In many safety filters the monitoring and intervention procedures are informed by
a safety-oriented control strategy that the filter considers as a potential \emph{fallback}.\footnote{In fact, as discussed in~\cref{sec:unified}, every safety filter can be viewed as relying on a fallback policy.}
This fallback safety strategy may be obtained by various means, including offline and runtime computation as well as manual engineering design.
Depending on the underpinnings of these components, a safety filter may offer different forms of guarantees, or none whatsoever.
\Cref{fig:safety_filter_block} shows a general diagram.

\begin{figure}
    \centering
    \includegraphics[width=1.23\textwidth]{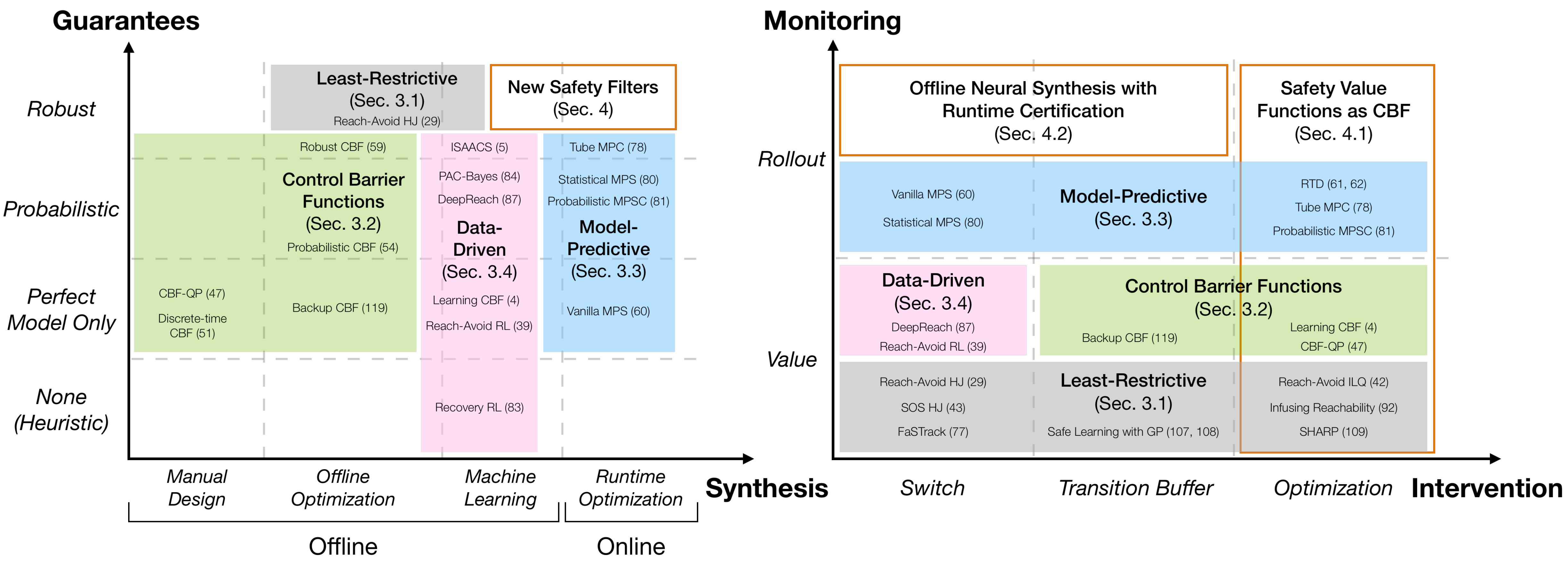}
    \caption{Summary of the safety filters reviewed in~\cref{sec:filters} and~\cref{sec:unified}. 
    The representative safety filter schemes are displayed under each category. \emph{Left:} Safety guarantees provided by safety filters vary according to the properties and synthesis of underlying fallback safety strategies. \emph{Right:} The runtime operation of safety filters is determined by the way they monitor the safety risk and intervene when they deem it to be unsafe.}
    \label{fig:safety_filter_summary}
\end{figure}

Safety filters can therefore be categorized attending to their safety monitoring methodology, their intervention scheme, the synthesis approach used to obtain the fallback safety strategy, and the safety assurances provided.
\cref{fig:safety_filter_summary} presents a visual summary of the main safety filter classes reviewed in this paper, organized by these four characteristics.

\subsection{Computing the Least-Restrictive Safety Filter: Dynamic Programming} \label{subsec:hj}
\begin{textbox}[h!]\section{Games of Kind and Games of Degree}
Rufus Isaacs, who worked alongside Richard Bellman at the RAND corporation in the 1950s and 60s,
pioneered the study of differential games
and
their connection with reachability and safety~\cite{isaacs1954differential}.
He drew a distinction between ``games of kind'', with a categorical outcome identifying a clear winner,
and ``games of degree'', in which players instead compete over a \emph{continuum} of possible outcomes.
Isaacs further showed that any game of kind can be represented implicitly as a game of degree.
For example, the game of whether a pursuer captures an evader can be transformed into a game in which the pursuer tries to minimize their closest future distance while the evader tries to maximize it.
If the separation falls below a specified capture radius, the pursuer wins the game of kind; otherwise, the evader wins. 
Importantly, the optimal strategies of the game of degree are also optimal for the corresponding game of kind:
minimizing the closest distance over time is an optimal approach to ensure capture, while maximizing it is an optimal way to escape.
This idea subsequently sparked numerous studies of pursuit-evasion games using level-set approaches~\cite{evans1984differential,mitchell2005timedependent,fisac2015reachavoid}.
\end{textbox}

A desirable safety filter is one that grants the robot as much freedom as possible to carry out its task while allowing no safety violations.
Computing the maximal safe set~$\safeSetMax$ \cite[also called \emph{viability kernel,}][]{aubin2011viability} is intrinsically
an optimal decision problem.
Under worst-case uncertainty, safety analysis can be viewed as a \emph{pursuit-evasion game} between a controller striving to 
avoid the failure set and an antagonistic disturbance agent aiming to steer the system into it~\citep{isaacs1954differential, lygeros1999controllers, valerii2001level}.
The Hamilton-Jacobi (HJ) reachability analysis framework uses 
level set representations
to transform the pursuit-evasion game, conceptually formulated as a \emph{game of kind}, into a \emph{game of degree}, by optimizing a continuous payoff \new{(see the aside on Games of Kind and Games of Degree)}.
Formally, in discrete time, the game’s solution can be obtained via the dynamic programming Isaacs equation~\cite{isaacs1954differential}:
\begin{marginnote}
    \entry{Margin function}{
    \looseness=-1
    a Lipschitz-continuous function $m\colon \reals^n\to\reals$
    encoding a given set $\mathcal{M} \subset \reals^n$ as its zero sublevel set,
    $\mathcal{M} = \{\state \mid m(\state) < 0\}$.
    A typical choice is the signed distance function to $\mathcal{M}$.}

    \entry{Signed distance function}{for any nonempty set $\mathcal{M} \subset \reals^n$, the signed distance function $\sgnDist{\mathcal{M}} \colon \reals^n \to \reals$ is defined as the closest distance to $\mathcal{M}$ for points $\state\not\in\mathcal{M}$, and the \textit{negative} closest distance to $\mathcal{M}^\compl$ for points $\state\in\mathcal{M}$,
    under some metric in $\reals^n$.}

    \entry{Viscosity solution}{a general solution concept~\cite{crandall1983viscosity} for HJ partial differential equations (PDE) that allows for discontinuities
    in the solution while satisfying the PDE boundary conditions almost everywhere.}
\end{marginnote}
\begin{equation}
    \valFunc_\tdisc (\state) = \min \left\{ 
        \consFunc(\state),\,
        \max_{\ctrl \in \cSet} \min_{\dstb \in \dSet} \valFunc_{\tdisc+1} \left( \dyn(\state, \ctrl, \dstb) \right)
    \right\},\,
    \tdisc \in \{0, 1, \cdots, \khorizon-1 \},~\valFunc_\khorizon (\state) = \consFunc(\state),
    \label{eq:isaacs_disc}
\end{equation}
where $\consFunc(\state)$ is a \emph{margin function} encoding the failure set $\failure$ as its zero sublevel set.
In the continuous time formulation, the outcome of the differential game is given by the viscosity solution to a Hamilton-Jacobi-Isaacs variational inequality that is the continuous-time counterpart of~\cref{eq:isaacs_disc}.
The safety \emph{value function} $\valFunc_\tdisc(\cdot)$ in~\cref{eq:isaacs_disc} can be computed
(up to numerical error) 
by backward induction in time, following the dynamic programming principle,
over a compact domain of the state space, in practice often represented by a numerical grid.
The induction converges
(typically in finitely many time steps)
to the infinite-horizon safety value function 
$\valFunc(\state) = \lim_{\tdisc \to -\infty} \valFunc_\tdisc (\state)$.
This value function encodes the \emph{maximal safe set} as well as an \emph{optimal safety policy}:
\begin{equation}
    \safeSetMax = \{\state \in \xSet \mid \valFunc(\state) \geq 0\}, \quad
    \policyCtrlOpt(\state) = \argmax_{\ctrl \in \cSet}\, \min_{\dstb \in \dSet} \valFunc \left( \dyn(\state, \ctrl, \dstb) \right).
    \label{eq:max_safe_set}
\end{equation}

The value function and policy in~\cref{eq:max_safe_set} can be used to construct a safety filter with value-based monitoring and switch-type intervention~\citep{chen2018signal}:
\begin{equation}
\label{eq:filter_hj}
    \safetyFilter \left(\state, \policyTask(\state) \right) := %
    \begin{cases}
        \policyTask(\state), \qquad & \min_{\dstb \in \dSet} \valFunc \Big( \dyn \left( \state, \policyTask(\state), \dstb \right) \Big) \geq 0 \\
        \policyCtrlOpt(\state), & \text{otherwise.}
    \end{cases}
\end{equation}
Intervention through $\policyCtrlOpt(\cdot)$ is only needed near the boundary of~$\safeSetMax$, to prevent $\valFunc(\state)$ from becoming negative.
While the filter is minimally restrictive in the sense of~\cref{prop:perfect_filter}, control chatter and erratic motion can result near the safe set boundary
due to intermittent overrides of any task policy that persistently attempts to leave~$\safeSetMax$.
This motivates
designing intervention strategies that yield
a smoother transition from performance-oriented to safety-preserving control, which we discuss further in~\cref{subsec:filter-aware}.

\looseness=-1
The \new{computation} complexity \new{and memory requirement} of numerically solving~\cref{eq:isaacs_disc} for general nonlinear dynamics
is exponential in the state dimension,
which limits its applicability to systems with no more than six continuous state variables.
To improve the tractability
of HJ methods, several approaches have been explored, including system decomposition~\citep{chen2018decomposition}, efficient initializations~\citep{herbert2019reachability}, and customized hardware design~\citep{bui2020realtime}.
Moreover, approximate solutions have been achieved through techniques such as deep reinforcement learning~\citep{fisac2019bridging, hsu2021safety}, deep supervised learning~\citep{rubiesroyo2019classification, bansal2021deepreach}, iterative linear-quadratic (ILQ) optimization~\citep{fridovichkeil2021approximate,nguyen2022back}, and sum-of-squares programming~\cite{singh2020robust}.
Rosolia and Borrelli~\cite{rosolia2017learning} proposed a runtime method to compute a gradually larger safe set by aggregating local model-predictive fallback policies.
Fridovich-Keil et al.~\cite{fridovich2019safely} developed a similar scheme to obtain an increasingly permissive safety filter for exploration of \emph{a priori} unknown environments,
bootstrapping the underlying reachability analysis and fallback policy with graph search methods.

Rather than aim to tractably approximate~$\safeSetMax$,
multiple approaches 
forego the maximal safe set and instead seek a smaller safe set $\safeSet \subset \safeSetMax$
that can produce a reliable, if more restrictive, safety filter.
\cref{subsec:cbf,subsec:rollout,subsec:critic} are dedicated to reviewing these methods.

\subsection{Value-Based Safety Filters: Control Barrier Functions} \label{subsec:cbf}

Initially developed by Wieland et al.~\cite{wieland2007constructive} and later refined by Ames et al.~\cite{ames2014control,ames2017cbf},
control barrier functions (CBFs) retain some important properties of the safety value function $\valFunc(\state)$\remove{, while no longer aspiring to or approximate}\new{. However, CBFs no longer encode or approximate} the maximal safe set~$\safeSetMax$.
The concept is motivated by earlier work on control Lyapunov functions~\cite{sontag1983lyapunov,sontag1989universal,artstein1983stabilization}, which---if found---provide a sufficient condition to stabilize a system.
An important feature of CBFs is that they allow a smooth intervention mechanism that relies on runtime control optimization.

Similar to least-restrictive filters,
a CBF safety filter $\safetyFilter(\state, \ctrl)$ renders a certain state function non-decreasing on its zero level set.
Formally, a (zeroing) CBF is a twice continuously differentiable function $\cbf:\reals^\nx\to\reals$ that encodes a safe set $\safeSet:=\{\state \colon \cbf(\state) \ge 0\}$ and,
additionally,
satisfies
the following inequality regarding its rate of decrease:
\begin{equation}
\label{eq:cbf_constr}
\sup_{\ctrl \in \cSet}  \nabla \cbf(\state)^\top \dynCont(\state, \ctrl) \ge -\alpha(\cbf(\state)),\quad \forall\state \in \safeSet\,,
\end{equation} 
\begin{marginnote}
    \entry{Class $\mathcal{K}$ function}
    {a continuous function $\alpha:\reals\to\reals$ is an extended class $\mathcal{K}$ function if it is strictly increasing with $\alpha(0)=0$. A simple choice of $\alpha(\cbf(\state))$ can be $\alpha(\cbf(\state)):=\cbf(\state)$.}
\end{marginnote}
{\!\!\!\!where}
$\alpha:\reals\to\reals$ is an extended class-$\mathcal{K}$ function, and $\dot{\state} = \dynCont(\state, \ctrl)$ are the system's \emph{continuous-time} dynamics.
If $\dynCont$ is control-affine, i.e. $\dynCont(\state, \ctrl) := \bar\dyn(\state) + \bar{g}(\state) \ctrl$, Inequality~\ref{eq:cbf_constr} becomes affine in control $\ctrl$, and a convex quadratic program (referred to as CBF-QP~\cite{ames2014control,ames2017cbf}) can be solved at each state $\state$ to produce a safe control $\safetyFilter(\state, \ctrl)$ that minimally deviates from a (potentially unsafe) task control $\policyTask(\state)$ while satisfying the CBF decrease rate condition:
\begin{subequations}
    \label{eq:CBF_QP}
    \begin{align}
        \safetyFilter \left( \state, \policyTask(\state) \right) = &  \argmin_{\ctrl \in \cSet} \ \frac{1}{2} \left\| \ctrl -  \policyTask(\state) \right\|^2 \\
        & \  \  \ \text{s.t.} \ \ \ \ 
        \nabla \cbf(\state) \dynCont(\state) + \nabla \cbf(\state) \bar{g}(\state) \ctrl + \alpha(\cbf(\state)) \geq 0\,.
    \end{align}
\end{subequations}

Since their debut, CBF-based safety filters have been extended to handle discrete-time problems~\citep{agrawal2017discrete}, perceptual uncertainty~\citep{dean2021guaranteeing, cosner2022self}, 
and, most recently,
systems with probabilistic and worst-case uncertainty~\cite{lyu2021probabilistic,alan2022safe}.
Contrasting with the efficient, systematic procedure for computing safe control actions given a known CBF,
\emph{finding} a valid CBF for a system of interest is often a nontrivial problem,
and the search for computationally tractable and data-efficient CBF synthesis methods remains an open research challenge.
Recent efforts have explored
analytical design~\cite{xu2017correctness}, sum-of-squares synthesis~\cite{wang2018permissive}, 
and learning from expert demonstrations~\cite{robey2020learning,lindemann2021learning,lindemann2021learningrobust}.
Unfortunately, existing CBF synthesis approaches tend to only work for restricted dynamical system classes and
suffer from scalability issues, especially faced with complex high-dimensional robotic systems in multi-agent environments.

Given the connection to safety value functions, it is not entirely surprising that CBF computation methods lose in scalability as they gain in generality.
In fact, when continuously differentiable, the safety value function $\valFunc$ is itself a valid CBF for safe set~$\safeSetMax$,
and can therefore
be used in a smooth CBF-type safety filter instead of a switch-type one.

\subsection{Rollout-Based Safety Filters} \label{subsec:rollout}

For high-dimensional dynamical systems, computing the optimal value function is prohibitive due to poor scalability,
while CBF methods lack general constructive mechanisms.
Instead, rollout-based safety filters aim to verify system safety at runtime by forward-simulating
(``rolling out'') 
the dynamics model or solving a trajectory optimization problem.

\subsubsection{Model predictive shielding (MPS)} \label{subsec:mps}

The basic model predictive shielding (MPS) algorithm~\cite{bastani2021safe-acc} works under the assumption that the dynamical model is fully accurate (no disturbance) and the filter has knowledge of a fallback policy $\fallback$ as well as a terminal safe set $\safeSet$ (typically small, e.g., at-rest states) that is invariant under $\fallback$.
Monitoring is conducted at each control cycle by simply checking if, after executing the candidate task control~$\policyTask(\state)$, 
the fallback policy $\fallback$ would safely drive the system state into $\safeSet$ within the prediction horizon (a sufficient condition for all-time safety, although not generally necessary, \new{and therefore typically conservative}).
Intervention is then a binary policy switch: if the safety monitor's check is successful, the filter lets the task control~$\policyTask(\state)$ through; if unsuccessful, it overrides it with the fallback policy~$\fallback$,
whose ability to preserve safety from $\state$ was necessarily verified at an earlier control cycle.
This simple scheme illustrates an important feature of safety filters:
they preserve (with the corresponding type of guarantee) the robot's future ability to maintain safety using a fallback safety strategy.
This results in a \emph{recursive safety} guarantee: as long as the fallback can maintain safety from the initial state, it will be able to maintain safety from every subsequent one.

While the methodology is agnostic to the choice of fallback policy~$\fallback$ and terminal safe set $\safeSet$, the conservativeness of the resulting filter depends greatly on them.
If poorly constructed, the fallback policy will need to be used often, in order to avoid reaching a state from which it cannot safely reach~$\safeSet$.
Similarly, a small terminal set~$\safeSet$ may be harder to reach by the fallback policy, triggering more frequent interventions.
On the opposite extreme, if given the maximal safe set~$\safeSet^*$ and an optimal safety policy~$\policy^*$, MPS recovers the least-restrictive filter of~\cref{eq:filter_hj} (since the safety monitor's validation only fails whenever the proposed action immediately drives the state out of $\safeSet^*$).

\subsubsection{Forward-reachable sets} \label{subsec:frs}
When the system \new{model} is \new{not perfectly accurate but} subject to bounded predictive uncertainty, its future safety can be checked using a \emph{forward-reachable set} (FRS).
Given an initial state and a control policy, the FRS contains all possible future states, including those reached under the worst-case uncertainty realization.
Safety is then determined in terms of whether the FRS \new{under the fallback policy eventually becomes \emph{fully contained} in the terminal safe set $\safeSet$ without previously \emph{intersecting} the failure set~$\failure$}.
\begin{marginnote}
    \entry{Forward-reachable set}
    {the set of all states $\state_\tdisc$ that might be reached by a system at time~$\tdisc$, given an initial state $\state_0$ and a control policy~$\policy$, under all possible $\dstb\in\dSet$.}
\end{marginnote}

Reachability-based Trajectory Design (RTD)~\citep{kousik2017safe,vaskov2019towards} considers \new{a} parameterized \new{space} 
\new{of nominal trajectories and associated tracking controllers}
and precomputes the FRS \new{of possible} tracking errors
\new{around each nominal trajectory}. 
At runtime, unsafe trajectories
\new{(those whose FRS cannot come to a full stop without colliding)}
are ruled out, 
and \new{task-driven control} optimization is 
\new{restricted to}
the \new{remaining} set of safe parameter values%
\new{---an optimization-type intervention scheme similar to the CBF constraint in~\cref{eq:CBF_QP}}.

Recent research efforts in computing FRS for verification include high-dimensional dynamics~\citep{kousik2020bridging}, hybrid systems~\citep{kochdumper2020reachability}, and imperfect observations~\citep{althoff2014online}.
We refer interested readers to a recent survey of set representations and computation for FRS~\cite {althoff2021set}.
Motivated by the growing use of neural network controllers,
the study of neural network output robustness~\citep{zhang2018efficient} has been extended to FRS computation for neural network--controlled systems~\citep{dutta2019reachability,hu2020reachsdp,everett2021reachability}.
Obtaining the FRS for a predefined \new{fallback} policy
decouples synthesis from verification, inducing a more tractable computational problem.
The downside is that the precomputed policy 
does not account for the FRS,
which can result in conservative filtering and diminished task performance.

\subsubsection{Tube MPC}
\new{Tube-based model predictive control (tube MPC) can be used to compute a local fallback policy via runtime optimization at each control cycle, rather than relying on a precomputed one.
}
Inspired by set and viability theory~\cite{blanchini1999set,aubin2011viability},
tube MPC optimizes a nominal trajectory alongside
a robust FRS (``tube'') encompassing all possible \new{tracking} deviations.
Existing techniques include an offline-synthesized tube with its linear auxiliary control policy~\cite{mayne2000constrained, mayne2005robust}, 
constraint tightening through contraction \new{theory} and sum-of-squares optimization~\cite{singh2017robust}, and online tube optimization with min-max differential inequalities~\cite{villanueva2017robust,hu2018real}.
A related approach is FaSTrack~\citep{chen2021fastrack},
which precomputes a tracking error bound and a nonlinear tracking policy by solving an offline pursuit-evasion game between a high-fidelity dynamical model and a lower-fidelity one used for runtime planning.

Tube MPC--based safety filters were first proposed by Wabersich et al.~\citep{wabersich2018linear} for linear dynamics with additive uncertainty.
At each control cycle,  
the tube MPC filter solves a runtime program
\new{to compute a nominal plan $\hat\state_\tdisc, \hat\ctrl_\tdisc$ with disturbance-free dynamics}:
\begin{subequations}
    \label{eq:tube_mpc}
    \begin{align}
        \min_{\hat\ctrl_\tdisc\in\hat\cSet} \;\; & \left\|\hat\ctrl_0 - \policyTask(x)\right\|^2
         \label{eq:mpc_obj} \\
         \st \quad & \hat\state_\tdisc \not\in\hat\failure,\quad\tdisc \in \{0,\dots,\khorizon-1\}\,,\\
                & \hat\state_\khorizon \in\hat\safeSet \, . \label{eq:mpc_tube}
    \end{align}
    where the adjusted failure set $\hat\failure\supset\failure$, terminal safe set $\hat\safeSet\subset\safeSet$, and control set $\hat\cSet\subset\cSet$ are computed to account for the tracking error and disturbance rejection effort under $\dstb\in\dSet$,
    ensuring that the entire FRS reaches $\safeSet$ clearing $\failure$.
    Monitoring is thus embedded in constraint feasibility, and the intervention is given by the computed solution as $\new{\safetyFilter} \left( \state, \policyTask(\state) \right) =  \hat\ctrl_0$.%
    \footnote{Similar to model-predictive shielding, whenever the constrained optimization cannot find a feasible solution, a fallback policy can be obtained
    from the last \new{feasible} cycle.
    In fact, these \new{prior feasible} controls are typically used as initialization to speed up the runtime optimization.}
\end{subequations}

Tube MPC--based safety filters have been extended to handle nonlinear dynamics by sampling~\citep{li2020robust}, statistical bounds~\citep{bastani2021safe}, and probabilistic reachable sets~\citep{wabersich2022probabilistic}.
Additionally, a data-driven dynamics model with state- and input-dependent uncertainty is considered~\citep{wabersich2021predictive}, and can be combined with more learning-based MPC methods.
In general, the optimization problem of tube MPC can be computationally intensive, 
especially for high-dimensional systems and complex constraints.
The next section reviews a variety of safety filters that aim to improve scalability by leveraging data and machine learning methods.

\subsection{Data-Driven Safety Filters} \label{subsec:critic}
Deep learning has proven to be an effective \new{framework} to 
\new{approximately solve}
optimal control problems \new{using} neural network--parameterized value functions (\emph{critics}) and control policies (\emph{actors}).
Recent work has used deep reinforcement learning (RL) with
a binary cost $\cost(\state, \ctrl) := \mathbbm{1}\{ \dyn(\state, \ctrl, 0) \in \failure \}$,
learning a safety critic that
can be seen as a risk monitor~\citep{srinivasan2020learning, thananjeyan2021recovery}.
This approach tends to exhibit slow convergence due to the sparse learning signal.
The fact that the algorithm can only learn by experiencing safety failures $\state\in\failure$ makes online learning impractical
for robotic applications where \new{high-fidelity} simulators are unavailable~\citep{hsuzen2022sim2lab2real}.
In some cases,
the safety critic can alternatively be trained offline from prior data, \new{in a \emph{post-mortem} fashion}~\citep{bharadhwaj2021conservative}.

Another line of research has 
pursued approximate solutions to the optimal safety problem encoded in \cref{eq:isaacs_disc}~\citep{rubiesroyo2019classification, bansal2021deepreach, fisac2019bridging, hsu2021safety, hsunguyen2023isaacs}.
Unlike sparse binary cost functions, the continuous payoff $\consFunc$ provides a dense training signal that additionally allows learning from near-failure events.
Self-supervised learning
has been employed to \new{approximately solve} the safety Bellman equation~\citep{rubiesroyo2019classification}, Isaacs equation~\citep{bansal2021deepreach,borquez2023parameterconditioned}, and give probabilistic guarantees~\citep{lin2023generating}.
These methods often assume that the algorithm has access to an explicit (uncertain) dynamics model and that the
min-max optimization in~\cref{eq:isaacs_disc} can be \new{solved efficiently at any given state $\state$}.
In contrast, deep RL approaches
only require that the system be available (physically or as a black-box simulator)
to collect state-control sequences.
Deep RL has \new{similarly} been used to approximate the solution to the safety Bellman equation~\citep{fisac2019bridging,hsu2021safety,li2022infinitehorizon} and the zero-sum Isaacs equation~\citep{hsunguyen2023isaacs}, 
as well as obtaining probabilistic guarantees~\citep{hsuzen2022sim2lab2real}.

Closely related to \new{approximating} the optimal \new{safety} value function, 
recent efforts aim to learn CBFs from data.
Robey et al.~\cite{robey2020learning} introduce an optimization-based method that synthesizes a 
CBF for nonlinear systems from expert demonstrations
\new{and verifies its validity through smoothness-based sufficient conditions}.
\new{Extensions handle}
handle hybrid systems~\cite{lindemann2021learning} and perception uncertainty~\cite{lindemann2021learningrobust}.
A conceptually similar approach~\cite{dawson2022safe} learns a neural-network-parametrized CBF based on a set of safe and unsafe state samples. 

\new{Data-driven} stability analysis can be used to \new{learn} controlled-invariant \emph{regions of attraction} for safety assurance.
For example,
Berkenkamp et al.~\cite{berkenkamp2017safe} propose a safe RL algorithm that yields high-probability 
stability guarantees during the learning process,
\new{using a Lyapunov-based fallback policy~$\policy$}
whose \new{learned} region of attraction is gradually enlarged as the system identifies its dynamics.
The \new{recent} stabilize-avoid formulation~\citep{so2023solving} 
aims to learn policies that safely drive the state to some equilibrium in a region of interest.

\begin{marginnote}
    \entry{Region of attraction}
    {the set of all initial states $\state$ from which trajectories converge asymptotically to a given equilibrium point $\state^*$, under a suitable policy~$\policy$.}
\end{marginnote}
Although neural network critics have shown practical utility in value-based monitoring, it must be kept in mind that the corresponding safety filters come with no guarantees. 
Using \remove{PAC-Bayes framework}
\new{statistical generalization theory, probabilistic}
guarantees can be provided under the assumption that training environments and the deployment conditions are drawn from the same (unknown) distribution~\citep{hsuzen2022sim2lab2real}.
In \new{\Cref{subsubsec:new_isaacs}}, we 
discuss an alternative to learned critics, which combines deep learning synthesis with model-predictive monitoring,
effectively treating 
the learned controller as an \emph{untrusted oracle} to guide the fallback strategy, which can then be verified at runtime through FRS rollouts to obtain \new{robust} safety guarantees.

\vspace{-3pt}
\subsection{Safety Filters for Multi-Agent Systems} \label{subsec:multi-agent}
Significant efforts have been directed toward the advancement of robotic systems capable of coexisting and engaging with other agents, particularly humans.
Enforcing safety becomes particularly challenging in interactive settings, as the robot and its \new{peer} agents may have coupled dynamics, limited communication capabilities, and conflicting interests.
Predominant safety methods for multi-agent systems assume worst-case agent behavior~\cite{leung2020infusing}, which can lead to the overly conservative decision-making known as the freezing robot problem~\cite{trautman2010unfreezing}, negatively impacting the robot's ability to complete assigned tasks. 
Several alternatives have been proposed with the aim of ``unfreezing'' the robot, to improve task performance without compromising on safety.

\subsubsection{Adaptive safety filters}
To prevent overly conservative robot behavior based on an \new{\emph{a priori}} worst-case assumption,
\emph{adaptive safety filters}
\new{dynamically} constrain the disturbance set $\dSet$, \new{modulating}
the range of admissible peer agent actions,
\new{based on environmental conditions or observed agent behavior}.
\emph{Forward reachability} methods~\cite{nakamura2023online, bajcsy2020robust} use
behavior prediction models in combination with reachability analysis to produce robust FRS predictions, which are then used for planning.
Althoff et al.~\cite{althoff2014online} impose modeling assumptions on driving behavior
\new{informed}
by traffic rules to yield less conservative forward-reachable sets of other road \new{users}.
Recent work~\cite{fisac2018probabilistically, fridovich2020confidence, tian2022safety}
dynamically adjusts the safety filter
\new{based on}
observed human behavior.
\new{While} these methods provide a tuning knob to balance safety and performance, \new{all-time} safety guarantees are usually lost 
\new{absent additional}
assumptions on human behavior.
\new{On the other hand, these approaches}
implicitly assume a \emph{static} understanding of the human's internal state during the entire safety analysis horizon,
\new{neglecting}
the robot's ability to gain information about the human \emph{in the future}.
This can be formally addressed in a safe partially observable stochastic game~\cite{hansen2004dynamic} (or dual control~\cite{feldbaum1960dual,mesbah2018stochastic}) framework, which is an emerging topic \new{showing some early research progress}~\cite{tian2021anytime, bonzanini2020safe, zhang2021occlusion, hu2023active,hu2023belgame}.

\subsubsection{Probabilistic safety filters}
Another commonly adopted safety method that relaxes the worst-case assumption relies on computing a \emph{probabilistically safe} control policy.
Fisac et al.~\cite{fisac2018probabilistically} propose to have the robot maintain a measure of its degree of confidence in a learned human behavior prediction model at runtime.
This allows the robot to plan probabilistically safe trajectories accounting for the (possibly varying) accuracy of human motion predictions.
Schildbach et al.~\cite{schildbach2015scenario} leverage scenario model predictive control~\cite{bernardini2011stabilizing} to compute safe lane change strategies for autonomous vehicles, accounting for the uncertain traffic environment, e.g., trajectories of other vehicles, encoded by a set of pre-generated scenarios.
Chance constraints~\cite{farina2016stochastic} are incorporated in the MPC problem, which establishes an upper bound on the probability of the ego vehicle violating the safety specifications.
Ultimately, however, under any such probabilistic approaches, safety can be compromised when other agents take actions 
\new{that are assigned low probability}
by the robot's prediction algorithm ~\cite[the ``long tail'' of unlikely events,][]{makansi2021exposing}.

\begin{figure*}[!hbtp]
  \centering
  \includegraphics[width=1.23\textwidth]{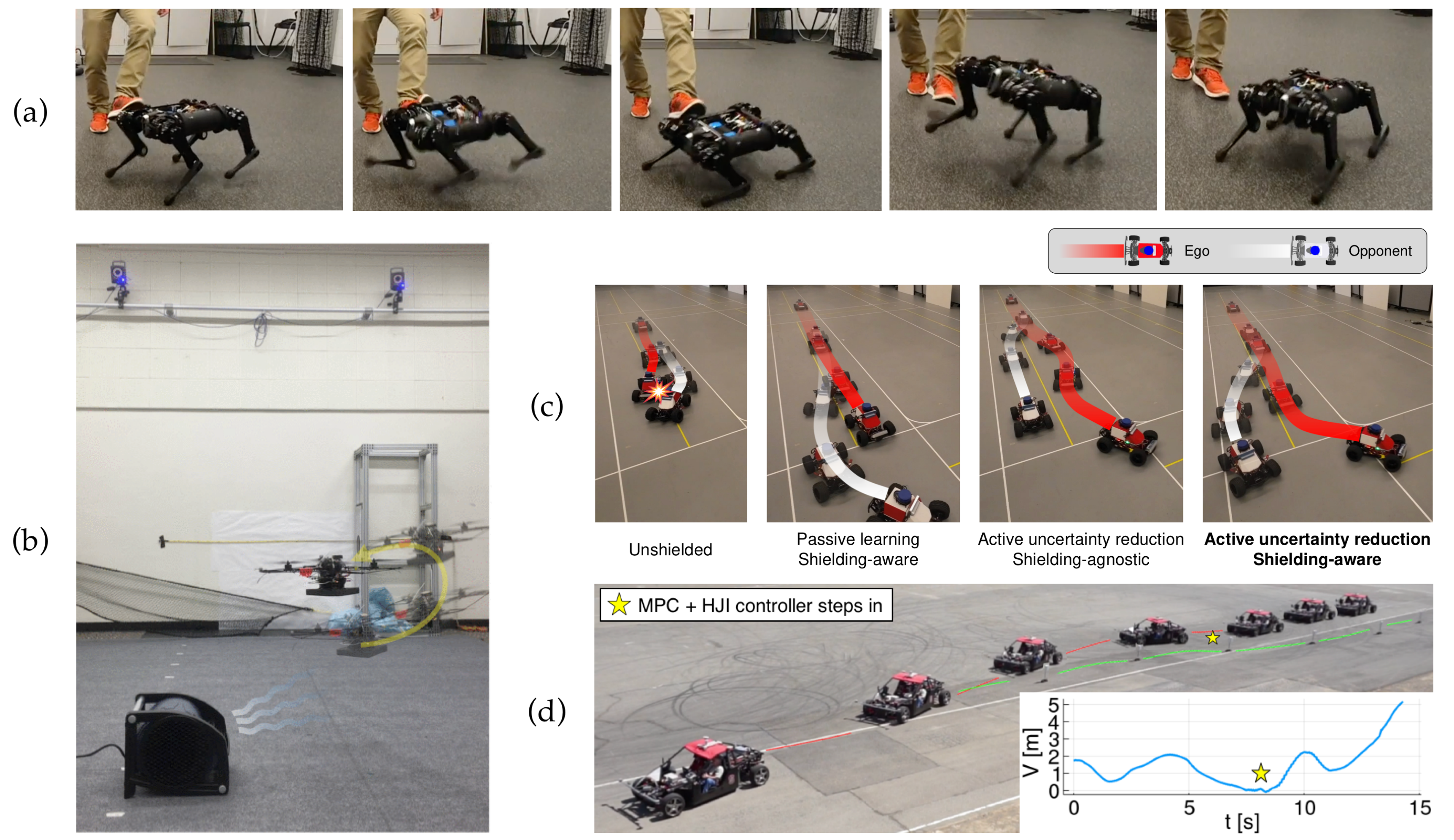}
  \caption{\label{fig:filter_aware}
      Applications of safety filters.
      \emph{(a)} A quadruped robot uses the adversarial RL policy~\cite{hsunguyen2023isaacs} as the safety filter to prevent falling when kicked by a human.
      \emph{(b)} A quadrotor learns to hover in the presence of wind disturbance while using an HJ-based fallback policy to ensure safety~\cite{akametalu2014reachability, fisac2019AGS}.
      \emph{(c)} A shielding-aware policy combined with a dual control scheme for safe active uncertainty reduction for autonomous driving~\cite{hu2022sharp, hu2023active}. The resulting trajectory is safe and chatter-free.
      \emph{(d)} A similar phenomenon was observed in Leung et al.~\cite{leung2020infusing}, in which an MPC planner is designed aware of an HJ-based safety filter, controlling the ego car to gently swerve to a nudging RC car (green trajectory).
      The value-based filter intervenes (indicated by the star) when $\valFunc(\cdot)$ defined in~\autoref{eq:max_safe_set} approaches $0$.
  }
\end{figure*}

\subsubsection{Filter-aware task policies} 
\label{subsec:filter-aware}
\new{Seeking a tighter integration between} safety and performance, 
some efforts have investigated endowing task policies with
\emph{safety filter awareness}.
\new{These approaches relax the safety-performance separation paradigm,
allowing the robot's task-driven decisions to explicitly account for the safety filter's behavior;
the goal is to avoid unnecessarily triggering interventions that may negatively impact performance.}
This idea was explored by Akametalu et al.~\cite{akametalu2014reachability} in the safe learning context, integrating the HJ safety value function $\valFunc$
into a the reinforcement learning reward signal, leading to a reduced frequency of safety filter interventions, which was shown to benefit the learning process.
More recently Leung et al.~\cite{leung2020infusing} studied filter awareness for safety-critical human-robot interaction in autonomous driving, with a HJ safety filter;
Hu et al.~\cite{hu2022sharp} developed a similar shielding-aware robust planning (SHARP) framework for a broader class of safety filters, and follow-up work~\cite{hu2023active} combined filter awareness with active uncertainty reduction to further mitigate the impact of safety overrides under low-probability human behaviors.

\subsection{Safety Filters for Imperfect Perception}  \label{subsec:perception_sf}
\new{The methods discussed thus far rely on the common assumption 
that the robot has reliable access to an accurate estimate of all relevant state variables.}
This assumption is
\new{increasingly challenged by modern} robotic applications.
Various forms of imperfect perception, ranging from occlusions to sensor failures can significantly endanger the robot's operation.
Under probabilistic uncertainty, \new{computing safe control strategies that account for} the robot's \emph{future} sensing ability can be formulated as a partially observable Markov decision process or stochastic game, which is generally intractable and requires approximation~\cite[see, e.g., ][]{brechtel2014probabilistic,bouton2018scalable,hubmann2019pomdp,sunberg2022improving};
the worst-case counterpart of this problem can be tackled using robust reachability analysis.
Koschi and Althoff~\cite{koschi2020set} propose an approach to compute the forward-reachable set of occluded objects, considering road structures and traffic rules in driving scenarios.
Laine et al.~\cite{laine2020eyes} provide a Hamilton-Jacobi safety analysis framework under the possibility that observations of \new{currently tracked} objects 
may be lost at a future time.
These methods conservatively model the robot's future access to information, relying on ``eyes-closed'' safety fallback strategies that do not respond to potential future observations.
\begin{marginnote}
    \entry{Information-space failure set}
    {set $\failure^\info$ of all information states $\info\in\infoSet$ that allow the possibility of some failure state,
    i.e., $\bar\xSet(\info) \cap \failure \neq \emptyset$.}
    \entry{Information-space safe set}
    {set $\safeSet\subseteq\infoSet$
    of information states from which some policy
    $\policy \colon \infoSet \to \cSet$
    maintains all-time safety
    for all realizations of the uncertainty.
    $\safeSet$ is controlled-invariant (under~$\policy$)
    and disjoint from the information-space failure set,
    i.e., $\safeSet \cap \failure^\info = \emptyset$.}
\end{marginnote}
In contrast, Zhang and Fisac~\cite{zhang2021occlusion}
incorporate the robot's measurement model into the safety synthesis to
obtain safety strategies that actively seek and adapt to informative observations of currently occluded regions and potential objects.
The problem is formulated as a two-mode pursuit-evasion game with an initially occluded pursuer (open-loop phase) that the robot may at some point discover within its field of view and subsequently avoid (closed-loop phase).
The framework provides a safety monitoring criterion that verifies whether the robot's strategy will enable it to detect any potential agent currently hidden beyond its field of view \emph{before} it becomes impossible to avoid a collision with it.
This leads to a simple switching safety filter with worst-case guarantees under imperfect observations.
Follow-up work~\cite{packer2023anyone}
plans occlusion-aware trajectories with a learned generative model that reasons about unobserved agents, who may be revealed in the near future, depending on the robot’s actions.
This learning-based approach enjoys better scalability at the expense of safety guarantees, which unfortunately are no longer available.
Control barrier functions 
(\cref{subsec:cbf})
have also been extended for robust motion planning against imperfect measurements~\cite{dean2021guaranteeing,lindemann2021learningrobust}. 
To conclude this section, we point out that safety analysis under imperfect state perception and uncertain agent intent (\cref{subsec:filter-aware}) share a common technical bottleneck, \new{namely} enabling the robot to strategically account for its \emph{future} ability to acquire new information.
Future research should focus on developing a \emph{single} safety analysis framework that unifies those two emerging fields.

\section{UNIFIED SAFETY FILTER THEORY} \label{sec:unified}

This section introduces a mathematical formalism that brings a broad class of the safety filters reviewed in this paper under a common lens.
The exposed modular structure can be used to reconceptualize a range of well-established safety filters, as well as to guide the construction of new ones.
The analysis focuses on \emph{robust} safety filters for uncertainty~${\dstb\in\dSet}$, while noting that the concepts and results have analogous probabilistic counterparts.

\begin{definition}[Safety Monitor]\label{def:monitor}
    Given an (uncertain) system as in~\cref{eq:dyn_disc_uncert,eq:obsrv} and an available fallback policy $\fallback \colon \infoSet \to \cSet$, a function $\monitor \colon \infoSet \times \cSet \to \reals$ is a safety monitor if it satisfies 
    the (one-way) implication
    \begin{equation}
        \monitor(\info, \ctrl) \ge 0 \implies
        \info\not\in\failure^\info \;\land\;
        \dyninfo(\info, \ctrl, \dstb) \in \safeSet^\shield,\; \forall \dstb \in \dSet,
    \end{equation}
    where
    $\failure^\info \subseteq\infoSet$ is the (possibly implicit) set of information states that do not preclude a current failure $\state\in\failure$, and
    $\safeSet^\shield\subseteq\infoSet$ is the (possibly implicit) set of all information states~$\info$ such that~$\fallback$ enforces~\cref{eq:safety_condition_robust} from all possible starting states $\state$ consistent with~$\info$.
\end{definition}

\begin{definition}[Safety Filter]\label{def:filter}
    A fallback policy~$\fallback \colon \infoSet \to \cSet$, safety monitor $\monitor \colon \infoSet \times \cSet \to \reals$, and intervention scheme $\safetyFilter \colon \infoSet \times \cSet \to \cSet$ constitute a (robust) safety filter if they jointly satisfy the implication:
        \begin{equation}
        \monitor\big(\info, \fallback(\info)\big) \ge 0 \implies \monitor\big(\info, \safetyFilter(\info, \ctrl) \big) \ge 0, \; \forall \ctrl \in \cSet \,.
    \end{equation}
\end{definition}

\Cref{def:monitor} establishes that a safety monitor only returns a positive value for a candidate control action if the known fallback policy is guaranteed to maintain safety \emph{after} the action is taken.
\Cref{def:filter} then requires that a (robust) safety filter intervene to preclude any control actions that fail the monitoring check whenever the fallback policy itself passes it.
The following general safety filter result then follows directly from the two definitions.

\begin{theorem}[Safety Filter]
\label{thm:general_sf}
Consider an (uncertain) system as in~\cref{eq:dyn_disc_uncert,eq:obsrv} and a safety filter as in~\cref{def:filter},
comprised by a fallback policy~$\fallback$, a safety monitor~$\monitor$ as in~\cref{def:monitor}, and an intervention scheme $\safetyFilter$.
If the system is deployed at
an initial information state ${\info_0 \in \infoSet}$
deemed safe by the safety monitor under the fallback policy, i.e., $\monitor \left( \info_0, \fallback(\info_0) \right) \ge 0$,
then its evolution under filtered dynamics $\dyn^\safetyFilter(\state,\ctrl,\dstb) := \dyn\big(\state,\safetyFilter(\info,\ctrl),\dstb\big)$ satisfies the safety condition in~\cref{eq:safety_condition_robust} for all possible task policies $\policyTask \colon \infoSet \to \cSet$.
\end{theorem}

The theorem's result relies on the fact that the safety filter preserves the positivity of the safety monitor's checks, and this in turn guarantees that it is always possible to robustly maintain all-time safety by resorting to the fallback policy.
The safety filter may never actually apply the fallback policy, but merely conserve the option to use it at a \emph{later} time.

We now showcase, \new{through a non-exhaustive set of example corollaries,} how safety guarantees for various classes of safety filters can be readily \new{rederived}
as special instances of~\cref{thm:general_sf}.
For fully observable settings, we simply let $\info \equiv \state$.

\begin{corollary}[Least-Restrictive Safety Filter]
    \label{cor:hj}
    The HJ safety filter with $\valFunc \colon \xSet \to \reals$, the maximal safe set $\safeSetMax\subset\xSet\subseteq\reals^\nx$ and intervention scheme $\safetyFilter$ defined in~\cref{eq:filter_hj} maintains all-time safety from any initially safe state $\state_0 \in \safeSetMax$.
\end{corollary}

\begin{proof}
    Let the safety monitor be $\monitor(\state,\ctrl):=\min_{\dstb \in \dSet} \valFunc \left( \dyn \left( \state, \ctrl, \dstb \right) \right)$
    and the fallback policy be $\fallback(\state):= \policyCtrlOpt(\state)$.
    The filter based on~\cref{eq:filter_hj} then satisfies~\cref{def:monitor,def:filter},
    and initially $\monitor(\state_0,\policyCtrlOpt(\state_0)\ge 0, \forall \state_0\in\safeSetMax$.
    Applying~\cref{thm:general_sf} completes the proof.
\end{proof}

\begin{corollary}[Control Barrier Functions]
    \label{cor:cbf}
    The CBF safety filter with CBF $\cbf \colon \xSet \to \reals$,
    with $\alpha(a) < a/\Delta \tcont, \forall a\in\reals$,
    safe set $\safeSet\subset\xSet\subseteq\reals^\nx$ and intervention scheme $\safetyFilter$
    defined as the CBF-QP in~\cref{eq:CBF_QP},
    maintains all-time safety from any initially safe state $\state_0 \in \safeSet$.
\end{corollary}

\begin{proof}
    Let the safety monitor be
    $\monitor(\state,\ctrl):=\min\{\cbf(\state), \dot{\cbf}(\state,\ctrl) + \alpha(\cbf(\state))\}$
    and the fallback policy (implicitly defined) be
    $\fallback(\state):=\arg\max_\ctrl \dot{\cbf}(\state,\ctrl)$.
    Indeed, if $\monitor(\state,\ctrl) > 0$, it must be that
    $\dot\cbf(\state,\ctrl) \ge - \alpha(\cbf(\state))$.
    Then, at the next control cycle,\footnote{Since the CBF formulation assumes continuous-time control, we neglect the integration error in $\cbf$ between control cycles.
    Practical CBF filter implementations typically enforce $\dot\cbf \ge \epsilon>0$.}
    $\cbf(\state') \ge \cbf(\state) - \alpha(\cbf(\state))\Delta\tcont > \cbf(\state) - \cbf(\state) = 0$,
    so $\state' \in \safeSet$, satisfying~\cref{def:monitor}.
    On the other hand, 
    if $\monitor(\state,\fallback(\state)) \ge 0$ this implies that the CBF-QP is feasible and therefore $\monitor(\state,\safetyFilter(\state,\ctrl)) \ge 0$ for any $\ctrl\in\cSet$,
    satisfying~\cref{def:filter}.
    Noting that $\monitor(\state_0,\fallback(\state_0)) \ge 0,\forall \state_0\in\safeSet$ and 
    applying~\cref{thm:general_sf} completes the proof.
\end{proof}

The next corollary combines model-predictive shielding (\cref{subsec:mps}) with forward-reachable sets (\cref{subsec:frs}). 
\begin{corollary}[Robust Model-Predictive Shielding]\label{cor:rollout}
    Consider a rollout safety filter with available fallback policy~$\fallback$, known terminal safe set $\safeSet$, and switch-type intervention scheme
    \begin{equation} \label{eq:rollout_intervention}
        \safetyFilter \left( \state, \ctrl \right) =
            \begin{cases}
                \ctrl, \quad &
                \new{\reach_\tplan} \cap \failure =\emptyset, \forall \tplan\in\{0,\dots,\khorizon\} \land
                \new{\reach_\khorizon} \subseteq \safeSet, \\
                \fallback(\state), & \text{otherwise},
            \end{cases}
    \end{equation}
    with $\new{\reach_0} = \{ \state \}$, $\new{\reach_1} = \{\dyn(\state, \ctrl,\dstb), \dstb\in\dSet\}$,
    and $\new{\reach_{\tplan+1}} = \{\dyn(\hat\state, \fallback(\hat\state),\dstb), \hat\state \in \reach_\tplan, \dstb\in\dSet\}, \tplan \ge 1$.
    This filter maintains all-time safety from any deployment state $\state_0 \in \safeSet$.
\end{corollary}
\begin{proof}
    Let $\monitor(\state,\ctrl) := \indicator\{ \new{\reach_\tplan} \cap \failure =\emptyset, \forall \tplan\in\{0,\dots,\khorizon\} \land \new{\reach_\khorizon} \subseteq \safeSet\}-\frac{1}{2}$
    be the safety monitor.
    Since the condition inside of the indicator $\indicator\{\cdot\}$ is sufficient for all-time safety ($\fallback$ drives all possible system trajectories safely into $\safeSet$, which is controlled-invariant),~\cref{def:monitor} is satisfied.
    Since the intervention scheme only applies $\ctrl$ when $\monitor(\state,\ctrl) \ge 0$ and otherwise overrides it with $\fallback(\state)$, the condition $\monitor(\state,\fallback(\state))\ge 0$ ensures that $\monitor(\state,\safetyFilter(\state,\ctrl))\ge 0$,
    meeting~\cref{def:filter}.
    Noting $\monitor(\state,\fallback(\state)) \ge 0, \forall\state\in\safeSet$, \cref{thm:general_sf} completes the proof.
\end{proof}

We now consider a safety filter for the imperfect-information setting of environment navigation with sensor occlusion, as discussed in~\cref{subsec:perception_sf}.
\begin{corollary}[Model-Predictive Safety Filter for Unmapped Environments]
    Consider a mobile robot with perfect estimation of its own state $\state^R$ and known discrete-time dynamics $\dyn^R$ navigating a static but initially unmapped environment $\mathcal{W}\subseteq\reals^3$. The robot is equipped with an onboard perception stack
     that conservatively identifies adjacent free regions and adds them to the known free space $\mathcal{W}^{\text{free}}(\info_\tdisc)\subseteq\mathcal{W}$ at each time $\tdisc$.
    The model-predictive safety filter
    that rejects any actions that would preclude the robot from coming to a full stop in $\khorizon$ steps within the currently known free space using a simple braking fallback policy $\fallback(\info)\equiv u^{\text{brake}}$, and overrides them with $u^{\text{brake}}$,
    maintains safety from any deployment information state $\info_0$ that has the robot at rest within the initially known free space.
\end{corollary}
\begin{proof}
    \looseness=-1
    In this case the failure set~$\failure^\info$ contains any information states $\info$ such that the robot's (known) position is outside of $\mathcal{W}^{\text{free}}(\info)$, and therefore \emph{may} be in collision.
    Let $\monitor(\info,\ctrl)$ return $+1$ if $\fallback$ brings the robot to a full stop from the next physical state $\dyn^R(\state^R,\ctrl)$ within another $\khorizon-1$ steps while remaining in the \emph{currently} known free space $\mathcal{W}^{\text{free}}(\info)$, and $-1$ otherwise.
    We can see that $\monitor(\info,\ctrl) = +1$ implies that the robot is currently inside $\mathcal{W}^{\text{free}}(\info)$, or equivalently $\info\not\in\failure^\info$;
    since additionally $\mathcal{W}^{\text{free}}(\info_{\tdisc+1})\supseteq\mathcal{W}^{\text{free}}(\info_{\tdisc})$ (the robot can still stop safely after any new free space is discovered),~\cref{def:monitor} is satisfied. The simple filter override with $u^{\text{brake}}$ satisfies~\cref{def:filter} by construction, and \cref{thm:general_sf} completes the proof.
\end{proof}

\subsection{Safety Value Functions as Control Barrier Functions} \label{subsubsec:new_cbvf}
Control barrier functions (CBFs) provide a smooth intervention mechanism
at the cost of settling for suboptimal safe sets and lacking general synthesis methods.
Exploiting the modular structure of safety filters,
a promising research direction is
to combine CBF-type intervention with other safety approaches that enjoy principled constructive \new{mechanisms}.
Ongoing research efforts explore the use of the HJ safety value function as a control barrier function.
Choi et al. bridge the dynamic programming safety formulation and the CBF condition to define and compute control barrier value functions~\citep{choi2021robust}.
Similar principles allow fine-tuning CBF candidates
online~\citep{tonkens2022refining}.
Most recently, Kumar et al.~\citep{kumar2023cbfddp} propose to locally approximate the safety value function by runtime differential dynamic programming, constructing an implicit CBF~\citep{chen2021backup} whose (also implicit) safe set approximates~$\safeSetMax$.

\subsection{Offline Neural Synthesis with Runtime Certification}\label{subsubsec:new_isaacs}
\looseness=-1
Model-predictive safety filters
(\cref{subsec:rollout}, \Cref{cor:rollout})
are designed to enforce safety using an \emph{arbitrary} fallback policy:
the more effective the fallback, the more freedom the filter can afford to grant.
This creates a natural synergy with modern neural safety policies trained by reinforcement learning (\cref{subsec:critic}), which can be effective in practice but typically lack guarantees.
Bastani and Li's statistical model predictive shielding~\citep{bastani2021safe}
uses reinforcement learning to train a fallback policy to reach a known terminal safe set.
By sampling a number of disturbance sequences for simulation, a statistical union bound is computed for probabilistic guarantees.
Hsu et al.~\citep{hsunguyen2023isaacs} propose co-training a safety controller and a worst-case disturbance policy
through adversarial reinforcement learning, approximately solving the Isaacs~\Cref{eq:isaacs_disc}.
Certification is then provided at runtime by robust rollout-based monitoring, at the cost of some added conservativeness incurred by the FRS computation.

\section{DISCUSSION} \label{sec:discussion}

Advancements in safe decision-making have been accelerating in recent years to meet the increasingly challenging demands of robot autonomy.
In this review, we have provided a unified perspective on a growing family of task-agnostic tools that \emph{filter} arbitrary control policies into safe ones,
exposing their shared structure and comparing their main features.
We conclude by briefly discussing
new research opportunities in safety-enforcing control.

\begin{issues}[FUTURE ISSUES]
\begin{enumerate}
    \item \textbf{Safety filters that account for runtime learning.}
    The design of safety filters
    for settings involving partial observability or agents with unknown intent
    usually 
    assumes a static information state, leading to overly conservative monitoring and intervention.
    Pushing the limit of safe autonomy in the information space calls for
    safety filters that tractably account for the system's runtime learning capabilities.
    \item \textbf{Safety filters that account for interaction.}
    Many safety filters assume non-responsive or actively adversarial behavior of other agents in the environment,
    to avoid precariously relying on the actions of others to ensure safety.
    Delineating practically acceptable assumptions for less conservative, yet no less meaningful safety assurances
    in interactive contexts like autonomous driving and robotic caregiving is likely to become a technical and regulatory priority in the coming years.
    \item \textbf{Safety filters with high-dimensional representations.}
    Modern robot autonomy frameworks increasingly rely on high-dimensional sensory input and learned latent representations.
    Reconciling these with the need for rigorous safety guarantees grounded in dynamical models is an open and important research challenge.
\end{enumerate}
\end{issues}

\section*{DISCLOSURE STATEMENT}
The authors are not aware of any affiliations, memberships, funding, or financial holdings that might be perceived as affecting the objectivity of this review.

\bibliography{references.bib}

\end{document}